\documentclass[floatfix, reprint, superscriptaddress, amsmath, amssymb, aps, prl]{revtex4-2}
\usepackage{graphicx}
\usepackage{amsmath}
\usepackage{amssymb}
\usepackage{amsfonts}
\usepackage{upgreek}
\usepackage{xcolor}
\usepackage[hidelinks]{hyperref}

\begin{document}

\title{Spin-strain coupling in nanodiamonds}
\author{Asad Awadallah}
\affiliation{Department of Chemical and Biological Physics, Weizmann Institute of Science, Rehovot 7610001, Israel}
\author{Inbar Zohar}
\affiliation{Department of Chemical and Biological Physics, Weizmann Institute of Science, Rehovot 7610001, Israel}
\author{Amit Finkler}
\affiliation{Department of Chemical and Biological Physics, Weizmann Institute of Science, Rehovot 7610001, Israel}

\begin{abstract}
    Fluorescent nanodiamonds have been used to a large extent in various biological systems due to their robust nature, inert properties and the relative ease of modifying their surface for attachment to different functional groups. Within a given batch, however, each nanodiamond is indistinguishable from its neighbors and so far one could only rely on fluorescence statistics for some global information about the ensemble. Here, we propose and measure the possibility of adding another layer of unique information, relying on the coupling between the strain in the nanodiamond and the spin degree-of-freedom in the nitrogen-vacancy center in diamond. We show that the large variance in axial and transverse strain can be encoded to an individual radio-frequency identity for a cluster of nanodiamonds. When using single nanodiamonds, this unique fingerprint can then be potentially tracked in real-time in, e.g., cells, as their size is compatible with metabolism intake. From a completely different aspect, in clusters of nanodiamonds, this can already now serve as a platform for anti-counterfeiting measures.
\end{abstract}

\maketitle

\section{Introduction}
There exist several types of site-labeled markers in biology and chemistry, and in a more general way as a means to provide a distinct signature or signal from a sample. Fluorescent markers are a good example of that, especially the extensive family of dyes \cite{Specht2017} but also nanoparticles \cite{Li2022}. Specifically, fluorescent nanodiamonds (FND) constitute by now an established technique \cite{Hsiao2016, Torelli2019} of marking specific organelles in cells by a tailored functionalization of the nanodiamonds' surface groups \cite{Chipaux2018}. Due to their robust and inert nature, FNDs are considered as highly stable emitters with a wide range of functional groups, applicable to many biological systems \cite{Krueger_2012, Wu_2015, Neburkova2017}.

Nevertheless, the fluorescence of one nanodiamond (for the same number of emitters per ND) is indistinguishable from that of another. That is to say, the number of photons emitted per second from one ND will be similar to that of another. It calls, then, for a class of such markers which one \textit{could} in fact tell them apart. Here we show, that by using NDs hosting nitrogen-vacancy (NV) centers, this feature comes about almost naturally, heralded by the spin-strain interaction \cite{Doherty2012, Doherty2013} of the NV with its host diamond matrix, providing an additional information layer to the fluorescence data, in a way akin to multicolor fluorescent dyes \cite{Sednev2015}. In this Letter we first describe the NV system Hamiltonian and introduce the relevant spin-strain terms to our experimental apparatus. We then present in detail a method to distinguish between different clusters of NDs dispersed on a substrate by measuring the effective (axial and transverse) strain components. This allows us to give an independent metric of characterization for the strain in a batch of such nanoparticles and, more importantly, provides a pathway for a generating a unique radio frequency identifier for this batch.

The latter comes about from the method by which we measure the strain, namely microwave manipulation of the NV spin state \cite{Jelezko2004a} and measurement by optical detection of magnetic resonance, or ODMR \cite{Gruber1997}.
\begin{figure}[hb]
    \includegraphics[width=0.45\textwidth]{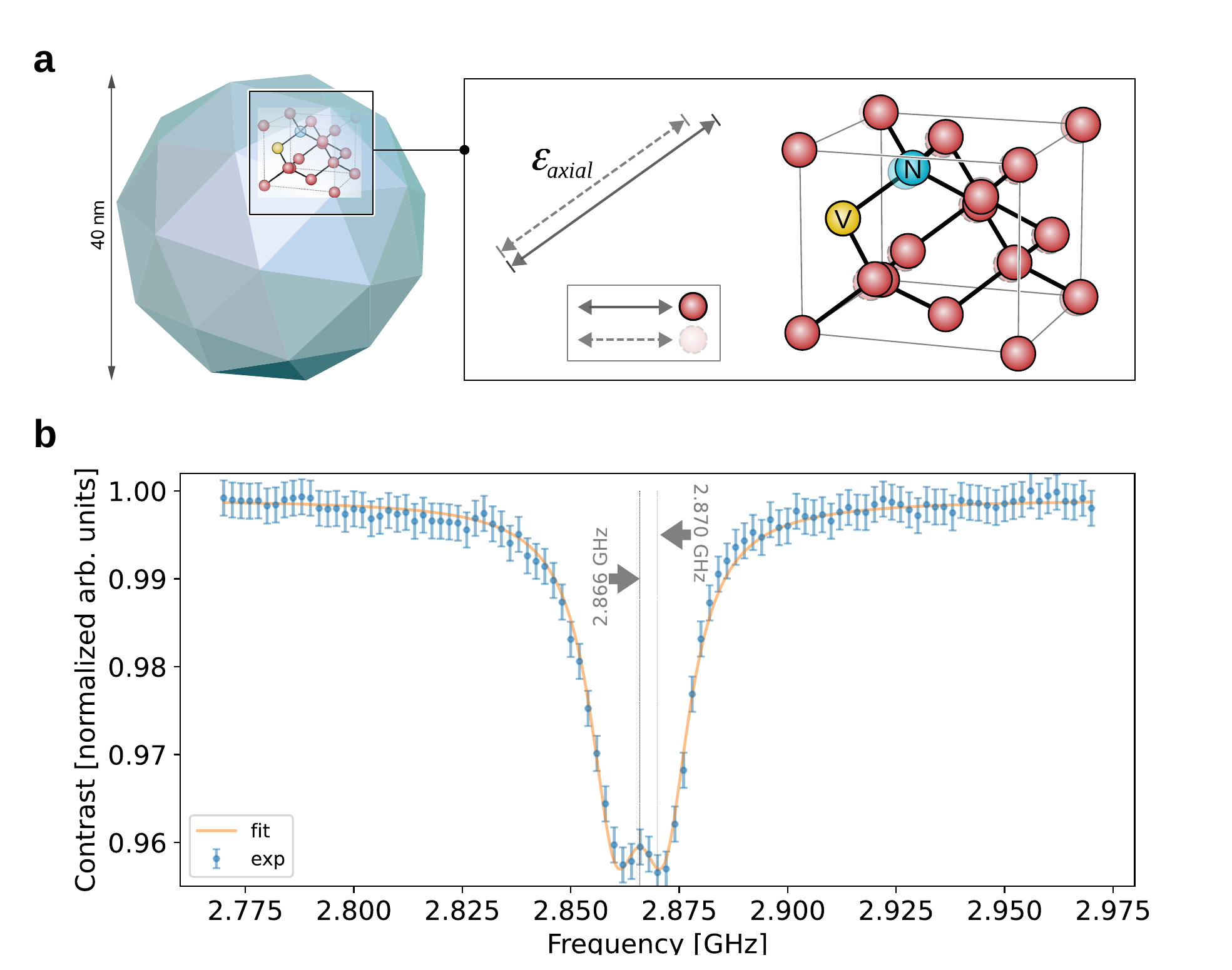}
    \caption{(a) A sketch of the diamond lattice with an N-V center, defining the direction of the strain field along (axial) the N$\rightarrow$V axis and perpendicular to (transverse) it as a result of pre-existing conditions due to milling. (b) A typical ODMR measurement from the nanodiamonds in the sample used in this paper. The two vertical lines show the axial strain leading to a shift from the unstrained zero-field splitting (in this case 4.1 MHz) and the splitting of the ODMR dip into two dips at zero magnetic field is a signature of the transverse strain (11.35 MHz).}
    \label{fig1}
\end{figure}
Strain in NVs has been studied extensively in the last decade, both experimentally \cite{Ovartchaiyapong2014, Teissier2014, Doherty2014} and theoretically \cite{Udvarhelyi2018, Barfuss2019}. Starting from the ground state Hamiltonian of the NV \cite{Doherty2012, Trusheim2016}, 
\begin{multline}
    \mathcal{H} = \left(D_\mathrm{gs}+\mathcal{E}_z\right)S_z^2 - \mathcal{E}_x\left(S_x^2 - S_y^2\right) + \mathcal{E}_y\left(S_xS_y+S_yS_x\right) \\ + g\mu_B \mathrm{S}\cdot\mathrm{B},
\end{multline}
we can calculate its eigenvalues, reaching an expression for the effect axial and transverse strain have on the transition frequencies of the NV in a given ODMR measurement, namely
$$
    \omega_{1,2} = D_\mathrm{gs} + \mathcal{E}_z \pm \sqrt{\mathcal{E}^2_\perp + \left(g\mu_B B_z\right)}.
$$
Here $\hbar$ was taken to be 1, $D_\mathrm{gs} = 2.87\,\mathrm{GHz}$ is the NV's zero-field splitting, $\mathcal{E}=\mathbf{d}_\mathrm{gs}(\mathbf{E}+\mathbf{\sigma})$ is the energy of interaction with electric and strain fields with $\mathbf{d}_\mathrm{gs}$ the electric dipole moment of the NV \cite{Trusheim2016}. Typical literature values for the axial ($\mathcal{E}_z(\sigma)$) and transverse ($\mathcal{E}_\perp(\sigma)$) change of interaction energy per unit strain, following Ref.\,\onlinecite{Trusheim2016}, are $\mathcal{E}_z(\sigma), \mathcal{E}_\perp(\sigma) = (20.60, 9.38)\,\mathrm{GHz}$. We should note that the values between the different reports \cite{Teissier2014, Ovartchaiyapong2014, Doherty2014} vary widely, yet are of the same order of magnitude. The axial strain direction is captured graphically in Fig.\,\ref{fig1}a, where the axial strain is defined to be along the N-V axis while the transverse one is perpendicular to it (not shown). While for many experiments, strain is a hindrance to be overcome or removed \cite{Appel2016, Kehayias2019}, we show how knowledge of its properties can be in fact a boon when using nanodiamonds.

\section{Results \& Discussion}
\begin{figure}[h]
    \includegraphics[width = 0.5\textwidth]{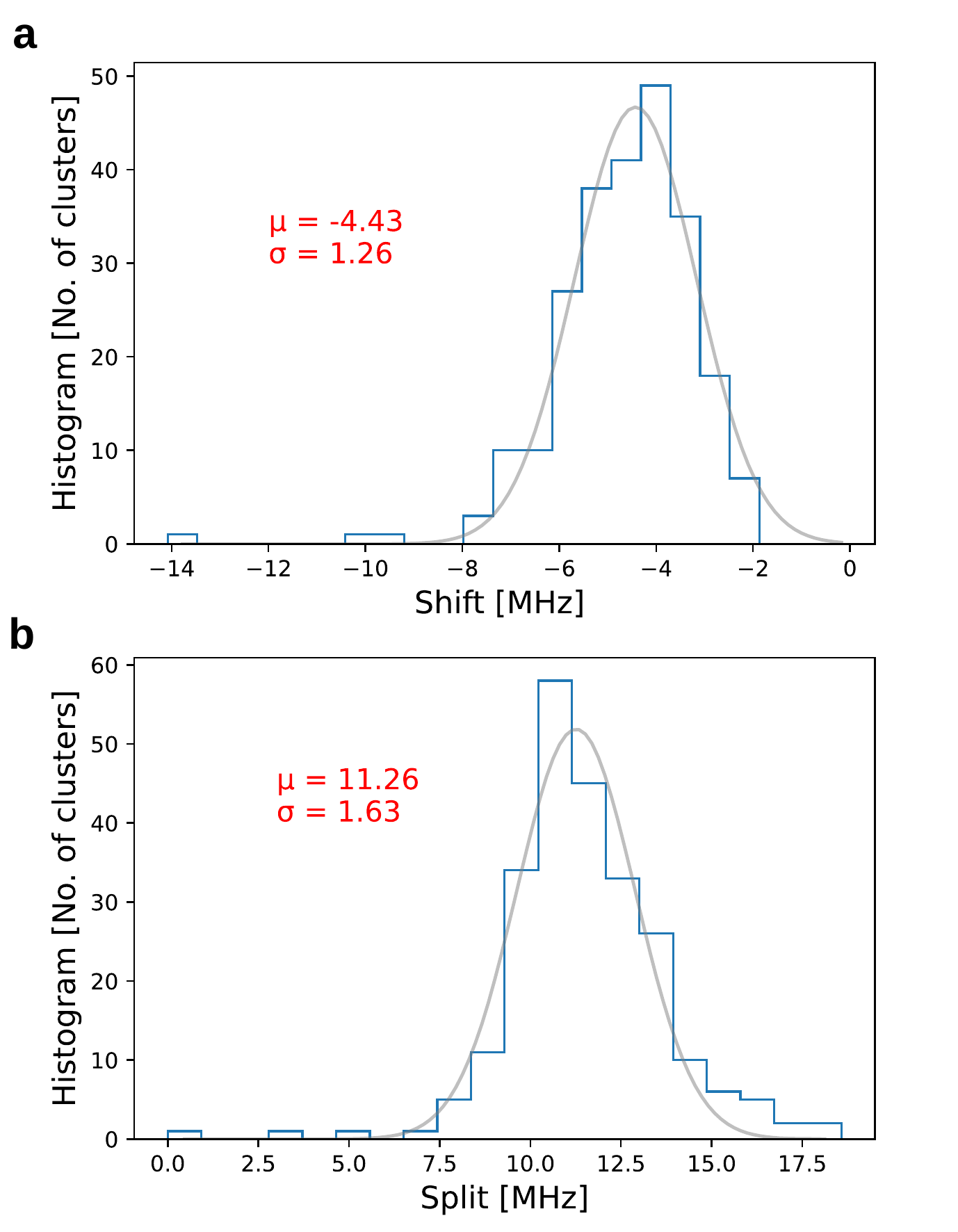}
    \caption{Histograms of the full cohort of NDs (247 in total) measured in this study. In each histogram, we also plotted the fit to a normal distribution (gray line). (a) Axial strain, with a mean value of -4.43 MHz, (b) Transverse strain, with a mean value of 11.26 MHz.}
    \label{fig2}
\end{figure}

A typical ODMR from the dispersed nanodiamonds (see Methods) is plotted in Fig.\,\ref{fig1}b, showing a shift from $D_\mathrm{gs}$ (axial strain) of $-4.10\pm 0.17$\,MHz and a split (transverse strain) of $11.35\pm 0.17$\,MHz (the Lorentzian linewidth, $\Gamma$, was $7.1\pm 0.2\,\mathrm{MHz}$). Following the central hypothesis of this work, we acquired ODMR spectra from a large cohort of points-of-interest (POI) in different regions-of-interest (ROI) on our sample. In Fig.\,\ref{fig2} 
we plot a histogram of both the frequency shift and split distribution among 247 different POIs. We obtain a mean shift of -4.43$\pm 1.26$ MHz and split of 11.26$\pm 1.63$ MHz. These, in turn, translate to a relative strain of 0.02\% in the axial direction and 0.12\% in the transverse direction, in agreement with prior reports on bulk diamonds \cite{Teissier2014}.

\begin{figure}[h]
    \includegraphics[width = 0.49\textwidth]{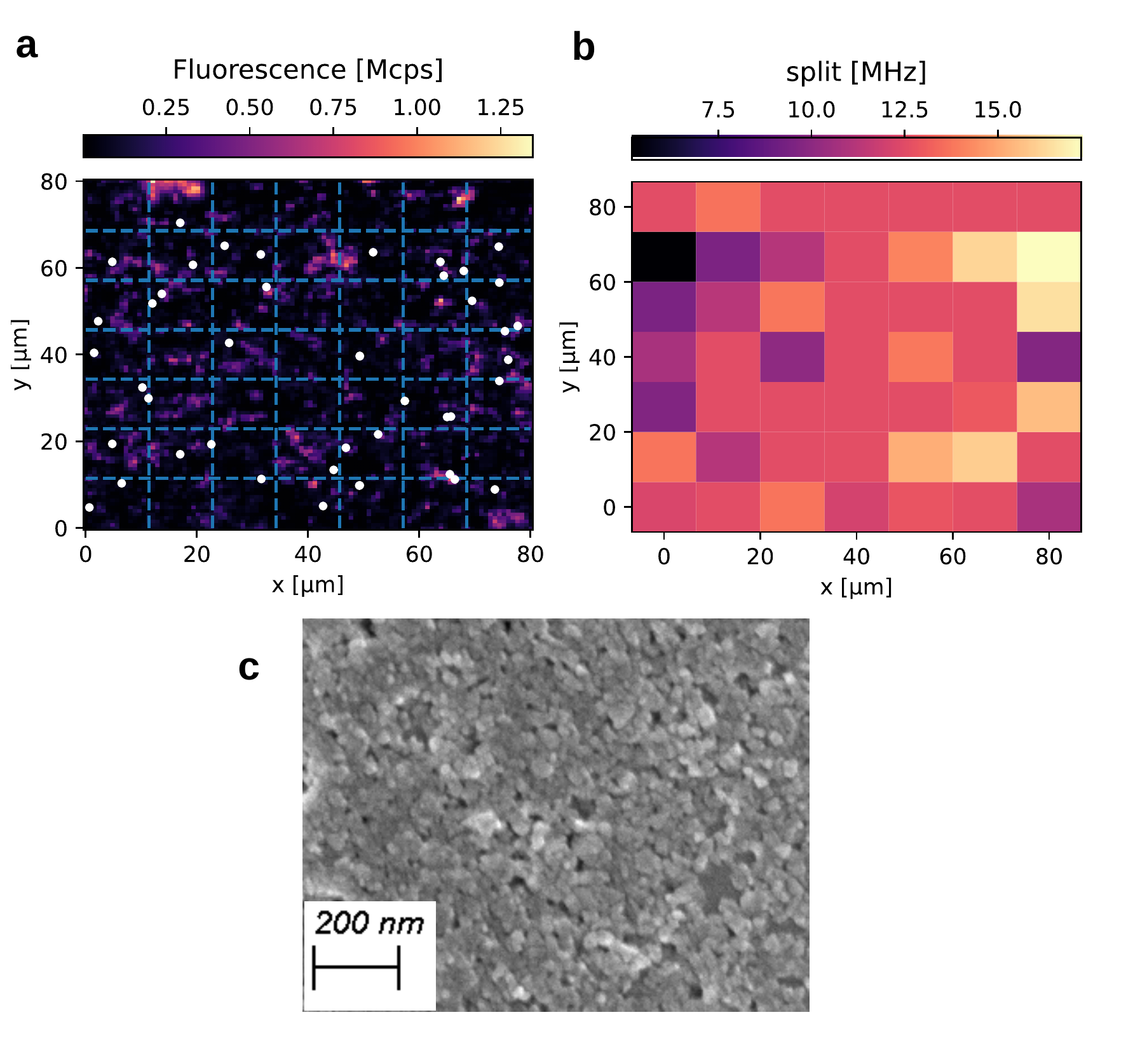}
    \caption{Visualization of a dispersion that is unique for each nanodiamond sample. The scan range, limited by the piezo scanner, is $80\times 80\,\upmu\mathrm{m}^2$. (a) A confocal scan with white dots denoting the positions of the NDs on which the ODMR measurements were performed; (b) A corresponding transverse strain map, where the $80\times 80$ image from (a) was sectioned into 7 rows and 7 columns, such that the different nanodiamonds were binned into one out of 49 pixels. (c) An SEM image of a typical ND dispersion, which corresponds to confocal images such as the one plotted in (a).}
    \label{fig3}
\end{figure}

These results suggest, that indeed, a collection of NDs dispersed on a glass substrate could be distinguishable from such a subsequent batch by correlating the strain parameters with the positions of the NDs. To visualize this idea, we plot in Fig.\,\ref{fig3}a an $80\times 80\,\upmu\mathrm{m}^2$ confocal map with white dots representing the positions of NDs at which we measured their ODMR spectra and extracted the strain parameters. 
The latter are plotted, corresponding to the (x,y) position in the confocal map, in Fig.\,\ref{fig3}b. All pixels with no data (no ND) were given the mean value of the strain for reasons of color offsetting. This last figure demonstrates the potential of this method: One can assign a unique two-dimensional color code (which can of course be translated to a standardized one such as the qrcode \cite{qrcode}, see SI), that can be verified against an existing database. Since the strain distribution is completely random, a large cohort will render the possibility of falsifying the reading unlikely. An additional parameter such as strain can add more security when applying such measures to anti-counterfeiting \cite{Hu2021}. Even more importantly, our results show that in combination with real-time ODMR tracking \cite{McGuinness_2011, Fujiwara2020}, one can unambiguously identify individual NDs in different environments. As we alluded to this in the introduction, one such prime example is the use of NDs inside living cells. Not only would one be able to monitor local temperature \cite{Kucsko2013}, pH~\cite{Rendler2017} and charge \cite{Petrakova2016} properties, but it would be also be possible to obtain this information with high confidence as to the identity and position of the sensor.

\begin{figure}
    \includegraphics[width = 0.49\textwidth]{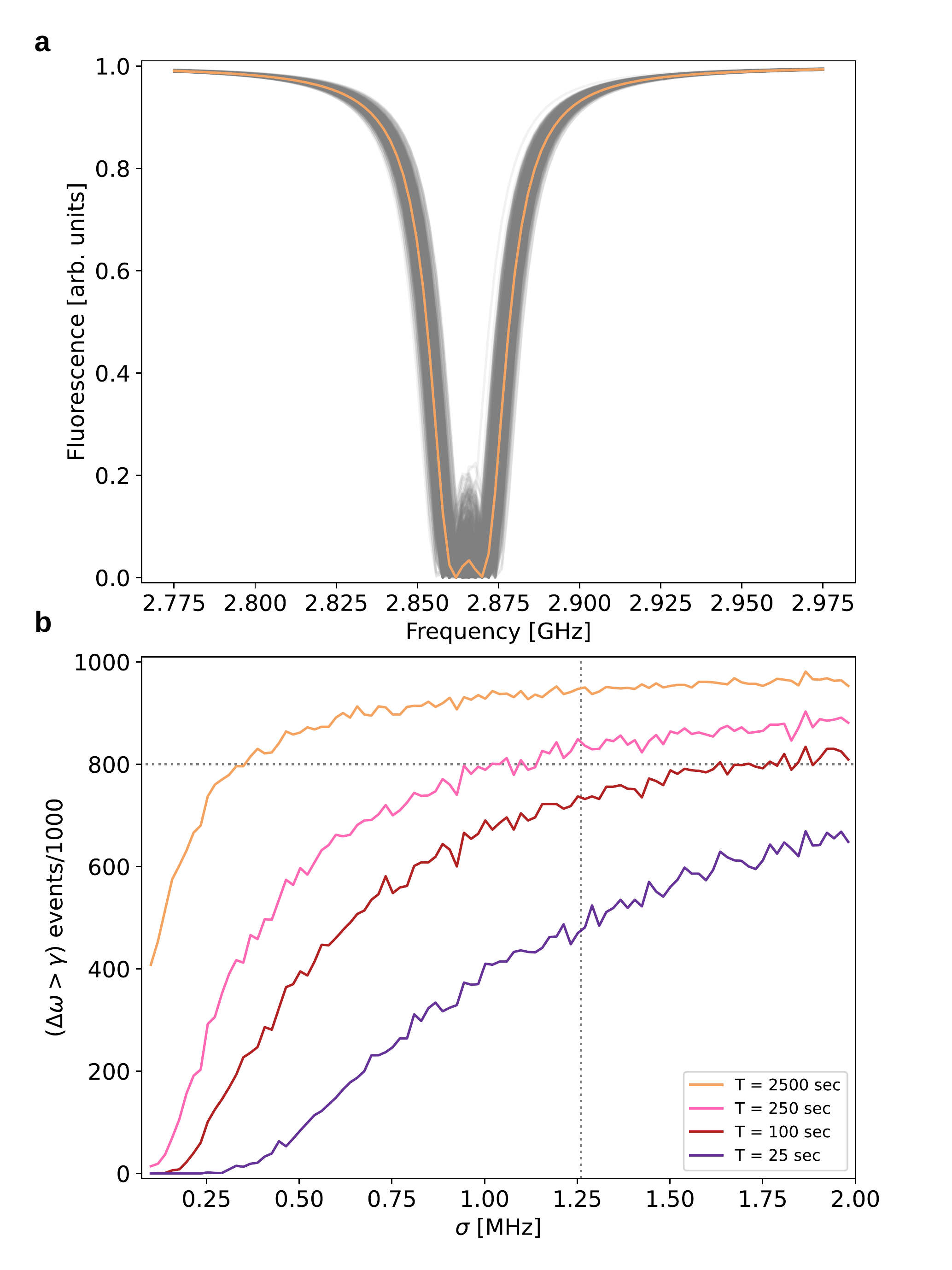}
    \caption{(a) Self-consistency check: Plotted is the sum of $n=800$ different ODMR curves with their shift and split parameters drawn from a normal distribution characterized by the mean and standard deviation obtained from Fig.\,\ref{fig2}. The gray curves are the individual ODMR plots. The orange curve, their sum, shows a clear split and shift from an unstrained ODMR curve; (b) Quantifying the ability to distinguish between two neighboring nanodiamond clusters. Shown are random draws of two frequencies based on the normal distribution of the measurements in Fig.\,\ref{fig2} for four different acquisition times (2500, 250, 100 and 25 seconds). The dashed lines provide a measure of fidelity, whereby one can estimate the needed physical parameters (measurement time, standard deviation of shift) to obtain a certainty higher than 80\% (chosen arbitrarily) of having the difference between the two frequencies larger than the standard deviation.}
    \label{fig4}
\end{figure}

\textbf{Quantifying ``uniqueness''}. The ODMR spectra that was used to calculate the mean axial and transverse strain was measured on a sample with a relatively high packing ratio (see Fig.\,\ref{fig3}c and also Methods \& SI), which indicates that an individual ODMR spectrum may include the contribution of several NDs and hence, NVs. Nine such representative ODMR spectra are plotted in the SI. To check the consistency of our conclusions, we plot in Fig.\,\ref{fig4} a calculation of an ODMR measurement which consists of $n=800$ different NVs, with their shift and splitting parameters drawn from a normal distribution function based on the mean and standard deviation values extracted above (and shown in Fig.\,\ref{fig2}). The number, $n$, was chosen so as to correspond to the approximate number of nanodiamonds in a confocal spot ($\sim 1\times 1\,\upmu\mathrm{m}^2$).
Due to the relatively narrow distribution of strain, we see that even a large cluster of NDs (at zero magnetic field) yields a distinct ODMR shift and split. We should note that the use of such clusters (or ensembles of NVs) is quite ubiquitous and in many cases makes it easier to acquire the signal \cite{Grant2022}. This allows us to now quantify how well one can distinguish between two such neighboring clusters, with their ODMR spectra modified by the shift and splitting randomly drawn from the normal distribution. For simplicity, we first consider just the frequency shift, and draw one thousand pairs of these clusters using the mean and standard deviation values from Fig.\,\ref{fig2}a. The first cluster's shift from 2.870\,GHz is denoted as $\omega_1$ and the second, $\omega_2$. We compare the difference between the two, $\Delta\omega = \left|\omega_1-\omega_2\right|$, to the standard deviation ($\gamma$) of our ODMR measurement for a given acquisition time. For example, $\gamma = 0.12$\,MHz for a total acquisition time of $T=2500$\,s. We label each of the one thousand draws with ``1'' if $\Delta\omega > \gamma$ (distinguishable) and ``0'' otherwise. We plot the draws for four different acquisition times (2500\,s, 250\,s, 100\,s and 25\,s) in Fig.\,\ref{fig4}b. For example, for a (arbitrary) fidelity threshold of 80\%, it is possible to tell apart neighboring clusters for a measurement time of 250\,s per ODMR if the standard deviation of the normal distribution is larger than $\sim 1.2\,\mathrm{MHz}$.
An immediate extension of this study would look at the spin-strain relation of individual NDs having on average a single NV. Moreover, we believe there should be a strong dependence of the mean value of both axial and transverse strain on the diameter of the ND, and accordingly, such a comparison would be beneficial for better understanding of both the statistical distribution itself and the origin of the strain. Finally, a high-temperature annealing process might change the strain distribution characteristics, and necessitates further study.

\section{Methods}
Nanodiamonds were purchased from Ad\'amas Nanotechnologies Inc.\,(model number NDNV40nmLw10ml). All measurements were done at zero applied magnetic field. A $\upmu$-metal shield (from Magnetic Shield Corporation) was placed around the sample, except for optical and microwave access, to minimize the effect of stray (and Earth's) magnetic fields.

Borosilicate cover slips (Epredia-Menzel-Gl\"aser 22 mm x 22 mm x 0.16 mm) were cleaned with acetone and isopropanol, then dried with dry nitrogen gas. This results in a relatively hydrophobic surface, leading to the aggregation of the nanodiamonds. Cleaning the cover slips with either Piranha Etch (mixture of H$_2$SO$_4$+H$_2$O$_2$) or burning them in an air environment at 500$^\circ$C leads to a hydrophilic surface (zero wetting angle) and a well-separated dispersion. In all the experiments described above, we used the hydrophobic surfaces.

The nanodiamonds were diluted from their original 1\,mg/ml concentration using type-1 ultrapure water (Milli-Q), and then drop-casted on a co-planar waveguide for the application of microwave tones. The measurements shown in Figs.\,2 and 3 are from a 20\,$\upmu\mathrm{g}$/ml concentration.

All measurements were performed on a custom-built confocal setup \cite{Zohar2022} with a QM-OPX orchestrating all pulse sequences and data acquisition. The entire experimental apparatus lies in a temperature- and humidity controlled room set to $23.0\pm 0.5^\circ$C and $35\pm 10\%$, respectively. Variations on the order of 0.5$^\circ$C would correspond to shifts of 35\,kHz in the zero-field splitting parameter \cite{Neumann2013}. Since this is approximately 1\% of the observed ODMR shifts and splitting, we can rule out the role of temperature as the generator of the observed ODMR shifts.

\paragraph{\textbf{Acknowledgements.}} We thank Petr C\'igler, Oliver Williams, Maabur Sow, Mateja Prslja and Olga Shenderova for helpful advice and consultation on dispersion of nanodiamonds. A.F.\,is the incumbent of the Elaine Blond Career Development Chair in Perpetuity and acknowledges support from the Israel Science Foundation (ISF grant 963/19) as well as the Abramson Family Center for Young Scientists and the Willner Family Leadership Institute for the Weizmann Institute of Science.

\bibliography{NVSPM}
\end{document}